\documentclass[aps,twocolumn,prc,superscriptaddress,altaffilletter,floatfix,preprintnumbers,amsmath,amssymb]{revtex4}

\usepackage{graphicx}% Include figure files
\usepackage{dcolumn}% Align table columns on decimal point
\usepackage{bm}% bold math
\usepackage{fancyhdr}
\usepackage{color}

\newcommand*{\dedx}{\ensuremath{dE/dx}}
\newcommand*{\ktopi}{\ensuremath{K/\pi}}
\newcommand*{\ptopi}{\ensuremath{(p + \bar{p})/\pi}}
\newcommand*{\roots}{\ensuremath{\sqrt{s_{_{NN}}}}}

\begin{document}

\pagestyle{fancy}
\fancyhf{}
\rfoot{\thepage}
\renewcommand{\headrulewidth}{0pt}

\title{Energy dependence of particle ratio fluctuations in central Pb+Pb collisions from 
$\sqrt{s_{_{NN}}} =$~6.3 to 17.3 GeV}

\affiliation{Nikhef, Amsterdam, Netherlands.}  
\affiliation{Department of Physics, University of Athens, Athens, Greece.}
\affiliation{Comenius University, Bratislava, Slovakia.}
\affiliation{KFKI Research Institute for Particle and Nuclear Physics,
             Budapest, Hungary.}
\affiliation{MIT, Cambridge, USA.}
\affiliation{Institute of Nuclear Physics, Cracow, Poland.}
\affiliation{Gesellschaft f\"{u}r Schwerionenforschung (GSI),
             Darmstadt, Germany.} 
\affiliation{Joint Institute for Nuclear Research, Dubna, Russia.}
\affiliation{Fachbereich Physik der Universit\"{a}t, Frankfurt,
             Germany.}
\affiliation{CERN, Geneva, Switzerland.}
\affiliation{Gesellschaft f\"{u}r Schwerionenforschung (GSI),
             Darmstadt, Germany.}
\affiliation{Institute of Physics \'Swi{\,e}tokrzyska Academy, Kielce,
             Poland.} 
\affiliation{Fachbereich Physik der Universit\"{a}t, Marburg, Germany.}
\affiliation{Max-Planck-Institut f\"{u}r Physik, Munich, Germany.}
\affiliation{Institute of Particle and Nuclear Physics, Charles
             University, Prague, Czech Republic.}
\affiliation{Department of Physics, Pusan National University, Pusan,
             Republic of Korea.} 
\affiliation{Nuclear Physics Laboratory, University of Washington,
             Seattle, WA, USA.} 
\affiliation{Atomic Physics Department, Sofia University St.~Kliment
             Ohridski, Sofia, Bulgaria.}
\affiliation{Institute for Nuclear Research and Nuclear Energy, Sofia,
             Bulgaria.}  
\affiliation{Institute for Nuclear Studies, Warsaw, Poland.}
\affiliation{Institute for Experimental Physics, University of Warsaw,
             Warsaw, Poland.} 
\affiliation{Rudjer Boskovic Institute, Zagreb, Croatia.}

%============================================================================

\author{C.~Alt}
\affiliation{Fachbereich Physik der Universit\"{a}t, Frankfurt,
             Germany.}
\author{T.~Anticic} 
\affiliation{Rudjer Boskovic Institute, Zagreb, Croatia.}
\author{B.~Baatar}
\affiliation{Joint Institute for Nuclear Research, Dubna, Russia.}
\author{D.~Barna}
\affiliation{KFKI Research Institute for Particle and Nuclear Physics,
             Budapest, Hungary.} 
\author{J.~Bartke}
\affiliation{Institute of Nuclear Physics, Cracow, Poland.}
\author{L.~Betev}
\affiliation{CERN, Geneva, Switzerland.}
\author{H.~Bia{\l}\-kowska} 
\affiliation{Institute for Nuclear Studies, Warsaw, Poland.}
\author{C.~Blume}
\affiliation{Fachbereich Physik der Universit\"{a}t, Frankfurt,
             Germany.}
\author{B.~Boimska}
\affiliation{Institute for Nuclear Studies, Warsaw, Poland.}
\author{M.~Botje}
\affiliation{Nikhef, Amsterdam, Netherlands.}
\author{J.~Bracinik}
\affiliation{Comenius University, Bratislava, Slovakia.}
\author{R.~Bramm}
\affiliation{Fachbereich Physik der Universit\"{a}t, Frankfurt,
             Germany.}
\author{P.~Bun\v{c}i\'{c}}
\affiliation{CERN, Geneva, Switzerland.}
\author{V.~Cerny}
\affiliation{Comenius University, Bratislava, Slovakia.}
\author{P.~Christakoglou}
\affiliation{Nikhef, Amsterdam, Netherlands.}
\author{O.~Chvala}
\affiliation{Institute of Particle and Nuclear Physics, Charles
             University, Prague, Czech Republic.} 
\author{J.G.~Cramer}
\affiliation{Nuclear Physics Laboratory, University of Washington,
             Seattle, WA, USA.} 
\author{P.~Csat\'{o}} 
\affiliation{KFKI Research Institute for Particle and Nuclear Physics,
             Budapest, Hungary.}
\author{P.~Dinkelaker}
\affiliation{Fachbereich Physik der Universit\"{a}t, Frankfurt,
             Germany.}
\author{V.~Eckardt}
\affiliation{Max-Planck-Institut f\"{u}r Physik, Munich, Germany.}
\author{D.~Flierl}
\affiliation{Fachbereich Physik der Universit\"{a}t, Frankfurt,
             Germany.} 
\author{Z.~Fodor}
\affiliation{KFKI Research Institute for Particle and Nuclear Physics,
             Budapest, Hungary.} 
\author{P.~Foka}
\affiliation{Gesellschaft f\"{u}r Schwerionenforschung (GSI),
             Darmstadt, Germany.} 
\author{V.~Friese}
\affiliation{Gesellschaft f\"{u}r Schwerionenforschung (GSI),
             Darmstadt, Germany.} 
\author{J.~G\'{a}l}
\affiliation{KFKI Research Institute for Particle and Nuclear Physics,
             Budapest, Hungary.} 
\author{M.~Ga\'zdzicki}
\affiliation{Fachbereich Physik der Universit\"{a}t, Frankfurt,
             Germany.}
\affiliation{Institute of Physics \'Swi{\,e}tokrzyska Academy, Kielce,
             Poland.}
\author{V.~Genchev}
\affiliation{Institute for Nuclear Research and Nuclear Energy, Sofia,
             Bulgaria.} 
\author{G.~Georgopoulos}
\affiliation{Department of Physics, University of Athens, Athens,
             Greece.}
\author{E.~G{\l}adysz}
\affiliation{Institute of Nuclear Physics, Cracow, Poland.}
\author{K.~Grebieszkow}
\affiliation{Institute for Experimental Physics, University of Warsaw,
             Warsaw, Poland.} 
\author{S.~Hegyi}
\affiliation{KFKI Research Institute for Particle and Nuclear Physics,
             Budapest, Hungary.} 
\author{C.~H\"{o}hne}
\affiliation{Gesellschaft f\"{u}r Schwerionenforschung (GSI),
             Darmstadt, Germany.}
\author{K.~Kadija}
\affiliation{Rudjer Boskovic Institute, Zagreb, Croatia.}
\author{A.~Karev}
\affiliation{Max-Planck-Institut f\"{u}r Physik, Munich, Germany.}
\author{M.~Kliemant}
\affiliation{Fachbereich Physik der Universit\"{a}t, Frankfurt,
             Germany.}
\author{S.~Kniege}
\affiliation{Fachbereich Physik der Universit\"{a}t, Frankfurt,
             Germany.}
\author{V.I.~Kolesnikov}
\affiliation{Joint Institute for Nuclear Research, Dubna, Russia.}
\author{E.~Kornas}
\affiliation{Institute of Nuclear Physics, Cracow, Poland.}
\author{R.~Korus}
\affiliation{Institute of Physics \'Swi{\,e}tokrzyska Academy, Kielce,
             Poland.} 
\author{M.~Kowalski}
\affiliation{Institute of Nuclear Physics, Cracow, Poland.}
\author{I.~Kraus}
\affiliation{Gesellschaft f\"{u}r Schwerionenforschung (GSI),
             Darmstadt, Germany.} 
\author{M.~Kreps}
\affiliation{Comenius University, Bratislava, Slovakia.}
\author{D.~Kresan}
\affiliation{Gesellschaft f\"{u}r Schwerionenforschung (GSI),
            Darmstadt, Germany.}
\author{M.~van~Leeuwen}
\affiliation{Nikhef, Amsterdam, Netherlands.}
\author{P.~L\'{e}vai}
\affiliation{KFKI Research Institute for Particle and Nuclear Physics,
             Budapest, Hungary.} 
\author{L.~Litov}
\affiliation{Atomic Physics Department, Sofia University St.~Kliment
             Ohridski, Sofia, Bulgaria.}
\author{B.~Lungwitz}
\affiliation{Fachbereich Physik der Universit\"{a}t, Frankfurt,
             Germany.}  
\author{M.~Makariev}
\affiliation{Atomic Physics Department, Sofia University St.~Kliment
             Ohridski, Sofia, Bulgaria.} 
\author{A.I.~Malakhov}
\affiliation{Joint Institute for Nuclear Research, Dubna, Russia.}
\author{M.~Mateev}
\affiliation{Atomic Physics Department, Sofia University St.~Kliment
             Ohridski, Sofia, Bulgaria.} 
\author{G.L.~Melkumov}
\affiliation{Joint Institute for Nuclear Research, Dubna, Russia.}
\author{A.~Mischke}
\affiliation{Nikhef, Amsterdam, Netherlands.}
\author{M.~Mitrovski}
\affiliation{Fachbereich Physik der Universit\"{a}t, Frankfurt,
             Germany.} 
\author{J.~Moln\'{a}r}
\affiliation{KFKI Research Institute for Particle and Nuclear Physics,
             Budapest, Hungary.} 
\author{St.~Mr\'owczy\'nski}
\affiliation{Institute of Physics \'Swi{\,e}tokrzyska Academy, Kielce,
             Poland.}
\author{V.~Nicolic}
\affiliation{Rudjer Boskovic Institute, Zagreb, Croatia.}
\author{G.~P\'{a}lla}
\affiliation{KFKI Research Institute for Particle and Nuclear Physics,
             Budapest, Hungary.} 
\author{A.D.~Panagiotou}
\affiliation{Department of Physics, University of Athens, Athens,
             Greece.} 
\author{D.~Panayotov}
\affiliation{Atomic Physics Department, Sofia University St.~Kliment
             Ohridski, Sofia, Bulgaria.} 
\author{A.~Petridis$^{\dagger ,}$}
\affiliation{Department of Physics, University of Athens, Athens,
             Greece.} 
\author{M.~Pikna}
\affiliation{Comenius University, Bratislava, Slovakia.}
\author{D.~Prindle}
\affiliation{Nuclear Physics Laboratory, University of Washington,
             Seattle, WA, USA.}
\author{F.~P\"{u}hlhofer}
\affiliation{Fachbereich Physik der Universit\"{a}t, Marburg, Germany.}
\author{R.~Renfordt}
\affiliation{Fachbereich Physik der Universit\"{a}t, Frankfurt,
             Germany.} 
\author{C.~Roland}
\affiliation{MIT, Cambridge, USA.}
\author{G.~Roland}
\affiliation{MIT, Cambridge, USA.}
\author{M.~Rybczy\'nski}
\affiliation{Institute of Physics \'Swi{\,e}tokrzyska Academy, Kielce,
             Poland.}
\author{A.~Rybicki}
\affiliation{Henryk Niewodniczanski Institute of Nuclear Physics, Polish Academy of Sciences, Cracow, Poland.\\ $^\dagger$deceased}
\author{A.~Sandoval}
\affiliation{Gesellschaft f\"{u}r Schwerionenforschung (GSI),
             Darmstadt, Germany.} 
\author{N.~Schmitz}
\affiliation{Max-Planck-Institut f\"{u}r Physik, Munich, Germany.}
\author{T.~Schuster}
\affiliation{Fachbereich Physik der Universit\"{a}t, Frankfurt,
             Germany.}
\author{P.~Seyboth}
\affiliation{Max-Planck-Institut f\"{u}r Physik, Munich, Germany.}
\author{F.~Sikl\'{e}r}
\affiliation{KFKI Research Institute for Particle and Nuclear Physics,
             Budapest, Hungary.} 
\author{B.~Sitar}
\affiliation{Comenius University, Bratislava, Slovakia.}
\author{E.~Skrzypczak}
\affiliation{Institute for Experimental Physics, University of Warsaw,
             Warsaw, Poland.} 
\author{G.~Stefanek}
\affiliation{Institute of Physics \'Swi{\,e}tokrzyska Academy, Kielce,
             Poland.}
\author{R.~Stock}
\affiliation{Fachbereich Physik der Universit\"{a}t, Frankfurt,
             Germany.}
\author{H.~Str\"{o}bele}
\affiliation{Fachbereich Physik der Universit\"{a}t, Frankfurt,
             Germany.}
\author{T.~Susa}
\affiliation{Rudjer Boskovic Institute, Zagreb, Croatia.}
\author{I.~Szentp\'{e}tery}
\affiliation{KFKI Research Institute for Particle and Nuclear Physics,
             Budapest, Hungary.} 
\author{J.~Sziklai}
\affiliation{KFKI Research Institute for Particle and Nuclear Physics,
             Budapest, Hungary.}
\author{P.~Szymanski}
\affiliation{CERN, Geneva, Switzerland.}
\affiliation{Institute for Nuclear Studies, Warsaw, Poland.}
\author{V.~Trubnikov}
\affiliation{Institute for Nuclear Studies, Warsaw, Poland.}
\author{D.~Varga}
\affiliation{KFKI Research Institute for Particle and Nuclear Physics,
             Budapest, Hungary.}
\affiliation{CERN, Geneva, Switzerland.} 
\author{M.~Vassiliou}
\affiliation{Department of Physics, University of Athens, Athens,
             Greece.}
\author{G.I.~Veres}
\affiliation{KFKI Research Institute for Particle and Nuclear Physics,
             Budapest, Hungary.} 
\affiliation{MIT, Cambridge, USA.}
\author{G.~Vesztergombi}
\affiliation{KFKI Research Institute for Particle and Nuclear Physics,
             Budapest, Hungary.}
\author{D.~Vrani\'{c}}
\affiliation{Gesellschaft f\"{u}r Schwerionenforschung (GSI),
             Darmstadt, Germany.} 
\author{A.~Wetzler}
\affiliation{Fachbereich Physik der Universit\"{a}t, Frankfurt,
             Germany.}
\author{Z.~W{\l}odarczyk}
\affiliation{Institute of Physics \'Swi{\,e}tokrzyska Academy, Kielce,
             Poland.}
\author{I.K.~Yoo}
\affiliation{Department of Physics, Pusan National University, Pusan,
             Republic of Korea.} 
\collaboration{The NA49 collaboration}
\noaffiliation

\date{\today}

\begin{abstract}
We present measurements of the energy dependence of event-by-event fluctuations in the 
\ktopi\ and \ptopi\  multiplicity ratios in 
heavy ion collisions at the CERN SPS. The particle ratio fluctuations were obtained for central Pb+Pb collisions at five 
collision energies, \roots, between 6.3 and 17.3~GeV.  After accounting for the effects of finite-number statistics
and detector resolution, we extract the strength of non-statistical fluctuations at each energy. For the 
\ktopi\ ratio, larger fluctuations than expected for independent particle production are found at all collision energies. 
The fluctuations in the \ptopi\ ratio are smaller than expectations from independent particle production, 
indicating correlated pion and proton production from resonance decays.
For both ratios, the deviation from purely statistical fluctuations shows an increase towards lower collision energies.
The results are compared to transport model calculations, 
which fail to describe the energy dependence of the \ktopi\ ratio fluctuations.
\end{abstract}

\maketitle

\section{Introduction}

Quantum Chromodynamics predicts that at sufficiently high temperature, strongly interacting matter will undergo a phase 
transition from hadronic matter to a state characterized by quark and gluon degrees of freedom, the quark-gluon plasma (QGP)
\cite{qgp}. Experimentally, strongly interacting matter under extreme conditions can be created in heavy ion collisions
at highly relativistic energies. Experiments have been performed over a very large range of center of mass 
collision energies, \roots, including $ 2.3~\mbox{GeV} < \sqrt{s_{_{NN}}} < 4.9$~GeV at the 
Brookhaven Alternating Gradient Synchrotron (AGS),  $ 6.3~\mbox{GeV} < \sqrt{s_{_{NN}}} < 17.3$~GeV 
at the  CERN Super Proton Synchrotron (SPS) and $ 19.6~\mbox{GeV} < \sqrt{s_{_{NN}}} < 200$~GeV at the Brookhaven Relativistic 
Heavy Ion Collider (RHIC).  

\begin{figure*}[tbh]
\begin{center}
\includegraphics[width=15cm]{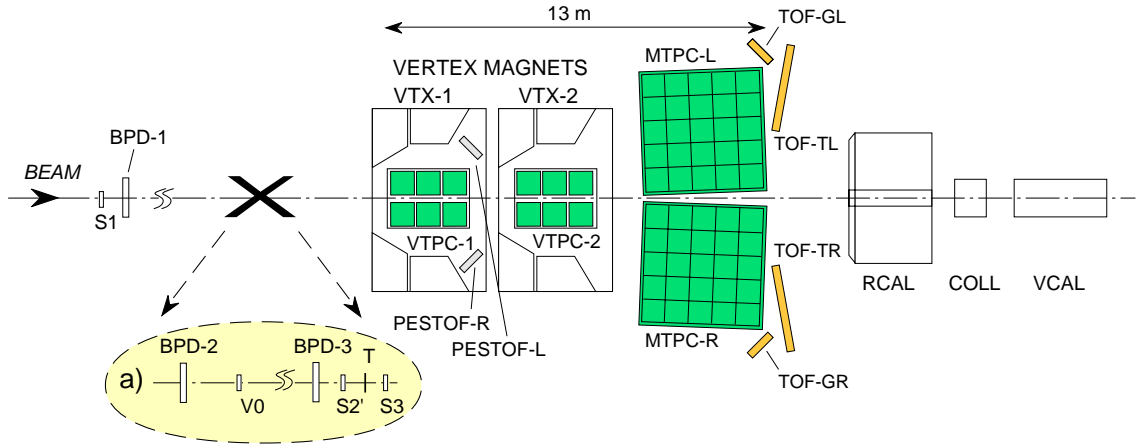} 
\end{center}
\vspace{-2.cm}
\caption{(Color online) The experimental set--up of the NA49 experiment
\protect\cite{na49_nim} and target arrangement (a) for Pb+Pb collisions. }
\label{na49_setup}
\end{figure*}

In the SPS energy range,  several hadronic observables in central Pb+Pb collisions
show qualitative changes in their energy dependence.
The ratio of average $\langle K^+ \rangle$ to $\langle \pi^+ \rangle$ yields 
exhibits non-monotonic behavior in the low-energy SPS range, 
close to $\sqrt{s_{_{NN}}} \approx 7.6$ GeV \cite{na49kpi}. 
In the same energy range, the slopes of the hadron
transverse momentum distributions show an approximately constant value, after a rapid rise at lower energies \cite{marekqm}.
These features are not observed in elementary interactions and appear to be unique characteristics of heavy ion collisions. 
Moreover, these observations have not been reproduced in hadronic transport models or statistical models of 
hadron production. The data are consistent with the expected signals of the onset of a phase transition
in heavy ion collisions at low SPS energies \cite{na49kpi,GaGo}. 
Recent measurements in Au+Au collisions at the highest available energies at RHIC indicate that in 
these collisions a highly collective, nearly thermalized system is formed \cite{rhic_whitepapers}, although the 
nature of the underlying degrees of freedom is still under debate.  A comparison of SPS to RHIC data shows
a rather smooth evolution of hadronic observables from the higher SPS energies up to the highest RHIC 
energies \cite{rhic_whitepapers,phobos_62GeV}.

Further information about the existence and nature of a phase transition in the SPS energy range can 
possibly be gained from studies of event-by-event fluctuations of final state 
hadron distributions.  Several mechanisms have been proposed that could lead to such event-by-event fluctuations,
including overheating-supercooling fluctuations due to a first order phase transition with large latent heat \cite{mishustin},
and fluctuations due to a phase of coexisting confined and deconfined matter (mixed phase) in 
varying relative abundances \cite{strange_hic}.  The presence or absence of such fluctuations might contain otherwise 
inaccessible information about the nature and order of a phase transition at these energies \cite{stock}. 

\begin{table*}[htb]
\begin{center}
\small
\caption{ \label{table1}
 Details of the data sets used in this analysis. For all data sets, the 3.5\% most central events
were selected. $y_{CMS}$ is the beam rapidity in the collision center-of-mass frame. 
The two numbers of tracks/event refer to the strict and loose track quality
cuts (see text).}
\begin{ruledtabular}
\begin{tabular}{c c c c c c}
 Beam energy (GeV) & \roots (GeV) & $y_{CMS}$ & Year taken & \# of events & Tracks/event\\
 \hline
20$A$  &  6.3  & 1.88 & 2002  & 140k  & 58 - 66   \\
30$A$  &  7.6  & 2.08 & 2002  & 170k  & 100 - 115 \\
40$A$  &  8.7  & 2.22 & 1999  & 160k  & 141 - 164 \\
80$A$  &  12.3 & 2.56 & 2000  & 140k  & 288 - 347 \\ 
158$A$ &  17.3 & 2.91 & 1996  & 120k  & 455 - 568 \\ 
\end{tabular}
\end{ruledtabular}
\end{center}
\end{table*}

Additional motivation for studies of fluctuations in kaon and proton production
comes from theoretical studies of the QCD phase diagram \cite{strange_hic,SMeSebye,StephCrit}.
It has been argued that the transition from hadronic to deconfined matter changes from a first order 
transition at large baryo-chemical potential, $\mu_B$, to a second-order transition or a cross-over at small 
$\mu_B$. This implies the existence of a critical endpoint at the end of the first order transition line. If 
freeze-out of a bulk strongly interacting system occurred close to the critical point, observable non-statistical
fluctuations in particle momentum distributions or baryon number correlations could be induced. While theoretical predictions
for the  location of the critical point and the strength of expected fluctuation signals are still under debate,
it is clear that a positive experimental signal for fluctuations related to the critical point 
could lead to major progress in our understanding of the QCD phase diagram.

To obtain experimental information directly related to the nature of the phase transition from 
fluctuations, systematic measurements in various regions of the $\mu_B$ versus temperature $T$  
phase diagram are necessary.  The energy density, temperature and baryo-chemical 
potential probed in the reaction can be controlled by varying the incident 
energy of the colliding system.  This will allow us to test if the observed structure 
in the energy dependence of event-averaged hadronic signatures 
is reflected in the collective phenomena probed by event-by-event fluctuations. 
Preliminary results by NA49 
showing a significant energy dependence of fluctuations in the kaon to pion and proton to pion 
multiplicity ratios were shown in \cite{Roland:2004pu}.  
In this paper, we present final results on the energy dependence of event-by-event fluctuations of 
these ratios in central Pb+Pb collisions 
over the SPS energy regime. 

Throughout the paper, we will define the event-by-event kaon to pion ratio as 
$ \ktopi \equiv  \frac{N{_{K^{+}}} + N_{K^{-}}}{N_{\pi^+} + N_{\pi^-}}$
and the event-by-event proton to pion ratio as 
$ \ptopi  \equiv  \frac{N_p + N_{\bar{p}}}{N_{\pi^+} + N_{\pi^-}}$.
Here, $N_{\pi^\pm}$, $N_{K^\pm}$, $N_{p}$  and $N_{\bar{p}}$ denote the observed pion, kaon, proton and anti-proton 
yields, respectively, within the acceptance for a given event.
It is important to note that all multiplicities, particle ratios and fluctuations reported in this paper 
refer to the observed yields in the experimental acceptance region. 
For the fluctuation measurement, an extrapolation to full phase space would 
only be possible using an assumption about the (unknown) nature of the underlying correlations or 
fluctuations. To allow a detailed comparison of theoretical calculations
to the results presented here, 3-dimensional acceptance tables for each collision energy are available online \cite{acc_tab_edms}.

\section{Experimental Setup and Data Selection}

The data presented here were taken with the NA49 experiment during runs from 1996 to 2002. 
The NA49 setup is shown schematically in 
Fig.~\ref{na49_setup} and described in detail in~\cite{na49_nim}.
Particle trajectories are measured using four large volume Time Projection Chambers (TPCs).
Two TPCs, VTPC-1 and VTPC-2, are placed in the magnetic field of two super-conducting dipole  magnets.
Two other TPCs, MTPC-L and MTPC-R, are positioned downstream of the magnets. The MTPCs were optimized
for a high precision measurement of the specific ionization energy loss \dedx\, which provides the main method 
of particle identification for this analysis. 
The target, a thin lead foil of about 0.01 interaction length for Pb ions, 
was positioned 80~cm upstream from VTPC-1.

\subsection{Event selection}
At all five beam energies the most central Pb+Pb collisions were selected 
based on the energy $E_{veto}$ deposited in a downstream calorimeter (VCAL) by projectile spectator nucleons.
The geometrical acceptance of the VCAL calorimeter was adjusted for each energy 
using a collimator (COLL)~\cite{na49_nim}. 

Details of the data sets used in this analysis are given in Table~1.
We only included the 3.5\% most central Pb+Pb collisions at each energy. The tight centrality selection was
used to minimize event-by-event fluctuations due to the residual dependence of the kaon to pion ratio
on system size and hence centrality. 
For the selected centrality range, we find that the remaining relative 
variation of the \ktopi\ and \ptopi\ ratios with $E_{veto}$ agrees within 1\% for the different collision 
energies.  In principle, this systematic variation could be washed out to varying degrees, 
depending on the resolution of the VCAL at the various beam energies. To constrain this 
effect, we used a Glauber calculation in combination with a parametrization of the $E_{veto}$ 
resolution of the VCAL calorimeter, taken from \cite{Alt:2006jr}.
With these simulations, we found the RMS variation in the number of participating nucleons 
in the selected collisions at the lowest energy to be 3.9\%, compared to 3.6\% at the highest 
energy.  As the absolute contribution of the \ktopi\ centrality dependence in the 3.5\% $E_{veto}$ bin
on the final dynamical fluctuation $\sigma_{dyn}$ is less than 1\%, the additional energy dependence 
induced by the varying $E_{veto}$ resolution is expected to be less than 0.1\%. No additional 
corrections are made for these contributions.

\subsection{\dedx\ particle identification}

The overall sensitivity of the \ktopi\ fluctuation measurements depends crucially on the resolution and stability 
of the event-by-event \dedx\ particle identification.
To optimize the stability of the \dedx\ measurement with respect to time,
variations of event multiplicity and possible background contributions, only the 
energy loss of the track in the MTPCs was used in this analysis. The tracking information from VTPC's was used
in the momentum determination and for rejecting background from secondary interactions and weak decays.

To eliminate a significant multiplicity dependence of the \dedx\ due to the high charge load
on the TPC readout chambers in central Pb+Pb events, a correction algorithm developed specifically
for the MTPCs was applied \cite{rolandc99}. The response function of the TPC amplifier/shaper
was determined by running an iterative shape fitting algorithm over the MTPC data. Using
the  response function obtained from this procedure, a channel-by-channel correction of the 
raw TPC charge measurements is performed. This unfolding procedure takes into account the charge history of
sets of neighboring channels, which are coupled via crosstalk effects
through the sense wires of each TPC readout chamber. 

These corrections
improve the average \dedx\ resolution by about 30\% from
$\sigma_{dE/dx}/\langle dE/dx \rangle = \mbox{5-6}\%$ to the final value of
$3.9\%$ for central Pb+Pb collisions. More importantly, the observed multiplicity
dependence of the \dedx\  measurement was reduced by more than 90\%, leaving
a change of less than $0.3\%$ over the selected multiplicity range. The values are 
quoted for the highest beam energy, corresponding to the 
highest multiplicities and charge loads in the TPCs. The performance of the 
algorithm was also cross-checked at each beam energy.

The MTPCs predominantly detect particles in the forward hemisphere of the collisions. This requires 
particle identification to be performed in the relativistic rise region, leading to a 
rather small separation in energy loss between different particle species. To allow 
for separation of kaons and protons, a cut in total momentum of $p_{tot} \geq 3$ GeV/$c$
is applied for all data sets.
Above this cut, the achieved \dedx\ resolution translates into an average separation of pions from 
kaons of about 2.1~$\sigma$ and of kaons from protons of about 1.8~$\sigma$.

\begin{figure}[t]
\includegraphics[width=8.5cm]{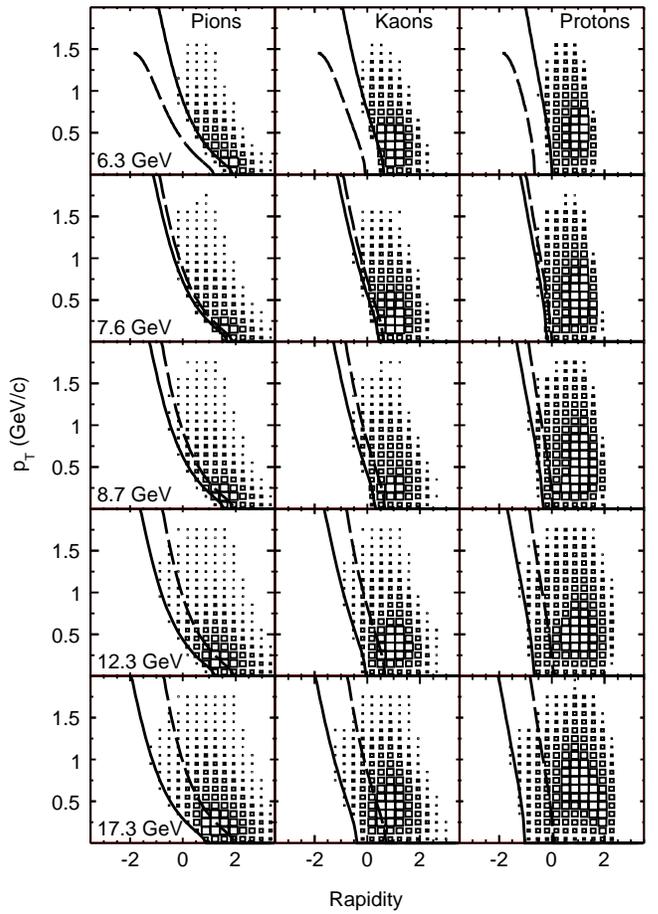}
\caption{\label{acc}Phase-space distributions of charged particles used for the fluctuation analysis 
at all collision energies, from $\roots\ \mbox {= 6.3}$ to 17.3~GeV. Data are shown using pion, kaon and proton
mass.  The solid line indicates the 3~GeV/c cutoff in total lab momentum. 
The dashed line indicates cutoffs at each energy that were used for 
the study of acceptance effects (see text). }
\end{figure}

In combination with the geometrical 
acceptance of the TPCs and the track quality cuts, the $p_{tot}$ cut determines the 
acceptance function for particles entering the analysis at each beam energy. To
minimize the rapidity acceptance shift for kaons, relative to CMS rapidity, 
the magnetic field settings for the two magnets were changed for each beam energy. Note that
in a fixed target setup it is not possible to simultaneously keep the relative rapidity acceptance constant for 
multiple particle species with different masses.  Figure \ref{acc} shows 
the phase space distributions of accepted particles at the different collision energies, 
assuming pion, kaon and proton mass for the left, center and right columns. 
For each plot, the solid line indicates the default $p_{tot} > 3$~GeV/c.
The figure illustrates the shifting acceptance, in particular for pions, 
in $(p_T,y)$ due to the changing acceptance and the constant PID cutoff in lab 
total momentum, $p_{tot}$. Shown also are dashed lines for alternative $p_{tot}$ cuts that were
applied to test effects of acceptance variations.

\section{Analysis Method}

\subsection{Event-by-event ratio determination}

\begin{figure*}[tb]
\includegraphics[width=17cm]{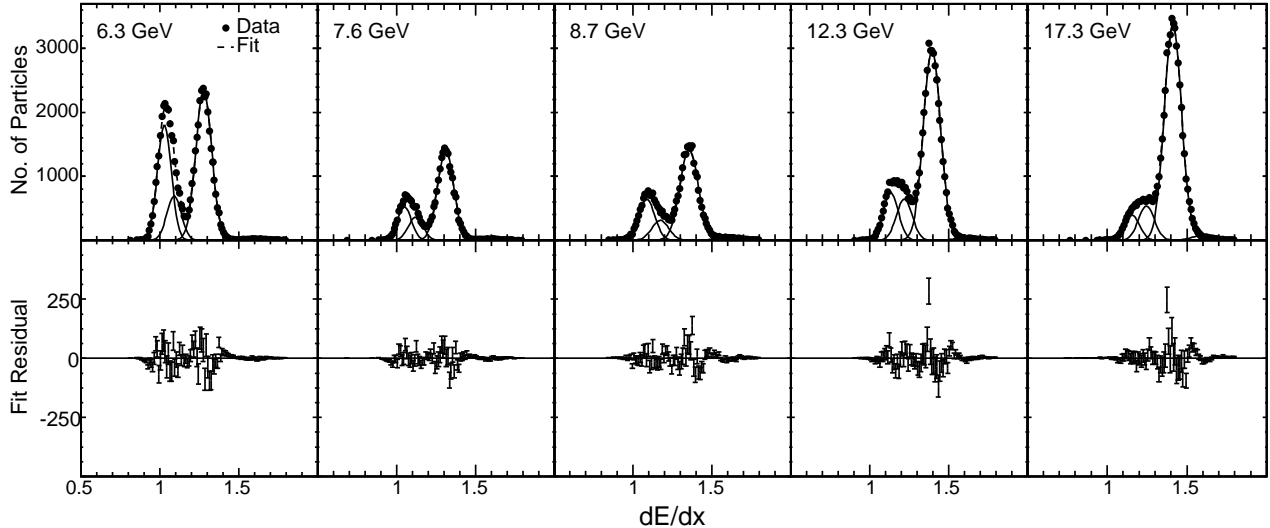}
\caption{\label{dedx} (Top) Event-averaged \dedx\ distributions (symbols) compared to
the extracted \dedx\ probability density functions (solid lines) 
for protons, kaons, pions and the sum of all particles. For all energies, data distributions and projected PDFs
are shown for a narrow momentum range ($\Delta p_{tot} \approx 1$~GeV/c) for $200~\mbox{MeV/c} < p_T <  400~\mbox{MeV/c}$, close to
$y = 0$ assuming kaon mass.
(Bottom) Fit residuals for the sum of \dedx\ probability density functions shown in the upper panels.}
\end{figure*}
%The bins:
%all energies
%200MeV < pT < 400MeV
%0 < phi < 1/4 pi
%charge = 1
%ptot bins:
%20GeV 4.3 - 5.2
%30GeV 5.2 - 6.3
%40GeV 7.6 - 9.1
%80GeV 9.1 - 10.9
%160GeV 13.2 - 15.9

Due to the finite number of particles detected for single Pb+Pb collisions and the limited particle
identification capabilities of a TPC in the relativistic rise region, it becomes crucial
to make maximum use of the available information when constructing an estimator
for the event-by-event particle ratios. The limited separation between different particle species
in the \dedx\ measurement does not permit a simple counting of particles in this experiment. Instead,
we use the unbinned distribution of particles in $(\vec{p},dE/dx)$ for each event to extract just two parameters,
\ktopi\ and \ptopi\, by performing an event-by-event  maximum likelihood fit \cite{oldk2pi, gazdzicki95}.

At each energy, the probability density functions (PDFs) for the particle momenta, normalized to
unity, $F_m(\vec{p})$, are determined for each particle species ($m = $ kaons, pions, protons, electrons).
We also evaluate the normalized PDFs for the truncated mean energy loss, $f_m (\vec{p};~dE/dx)$, as a
function of particle momentum for each species.

For the event-by-event fit, the relative yield of different particle species is characterized by
parameters $\Theta_{m}$, such that $\sum_m \Theta_{m} = 1$. 
These parameters are determined for each
event by maximizing the following likelihood function:
\begin{equation}
%L = \prod_{i=1}^n(P(p,dE/dx|\Theta)
L = \prod_{i=1}^n [\sum_m \Theta_{m} F_{m} (\vec{p}_i) f_{m} (\vec{p}_i;(dE/dx)_i)].
\end{equation}
Here $\vec{p}_i$ denotes the observed momentum vector and $(dE/dx)_i$ the specific energy loss for each particle $i$ in
the event.  The parameters $\Theta_{m}$ were constrained to be positive for the maximum likelihood fit.
The maximum likelihood fit then yields directly the event-by-event \ktopi\ and \ptopi\
particle ratios. No other parameters are fitted event-by-event.

The PDF used in the fit are determined from event-averaged data for each beam energy. The particles from each
sample are split into bins in total momentum $p_{tot}$, transverse momentum $p_T$, azimuthal angle $\phi$ and charge.
The \dedx\ distribution of the particles in each bin is fitted by four Gaussian distributions, one for each particle
species. The fit results for all bins are stored, providing the description of the PDF for each bin. 
The position and width of the Gaussian distributions describe the \dedx\ PDF, $f_m$, 
while the integrals of the distributions in the momentum bins describe the PDF of the
momentum distributions, $F_m$.

The top row in Figure \ref{dedx} shows the event-averaged \dedx\ distributions for positive particles in
narrow momentum slices $\Delta p_{tot} \approx 1$~GeV/c for each of the 5 collision energies.
For each energy, the selected region was chosen 
close to rapidity $y=0$, assuming kaon mass, with transverse momenta $200~\mbox{MeV/c} < p_T <  400~\mbox{MeV/c}$. 
Also shown in the top row as solid lines are the extracted PDFs for $p$, $K$, $\pi^+$ and $e^+$ in the selected momentum slice
and the PDF sum. The bottom plots show the residual difference between data and sum PDF, illustrating the 
quality of the \dedx\ parametrization. A small, systematic structure in the residuals can be observed
at all energies, resulting from small non-Gaussian tails in the \dedx\ distributions. The dominant 
contribution to these tails comes from a variation in the number of hits in individual particle tracks, 
where tracks with a smaller number of hits lead to a wider \dedx\ distribution. We did not include 
these non-Gaussian tails in the \dedx\ PDFs. 
The resulting effect on the extracted fluctuations was tested by changing the \dedx\ scaling of the width of the \dedx\ fits (see below), which changes the relative contribution of the non-Gaussian tails, and by varying track quality cuts, and is included in the final systematic uncertainties.

\begin{figure}[t]
\includegraphics[width=9cm]{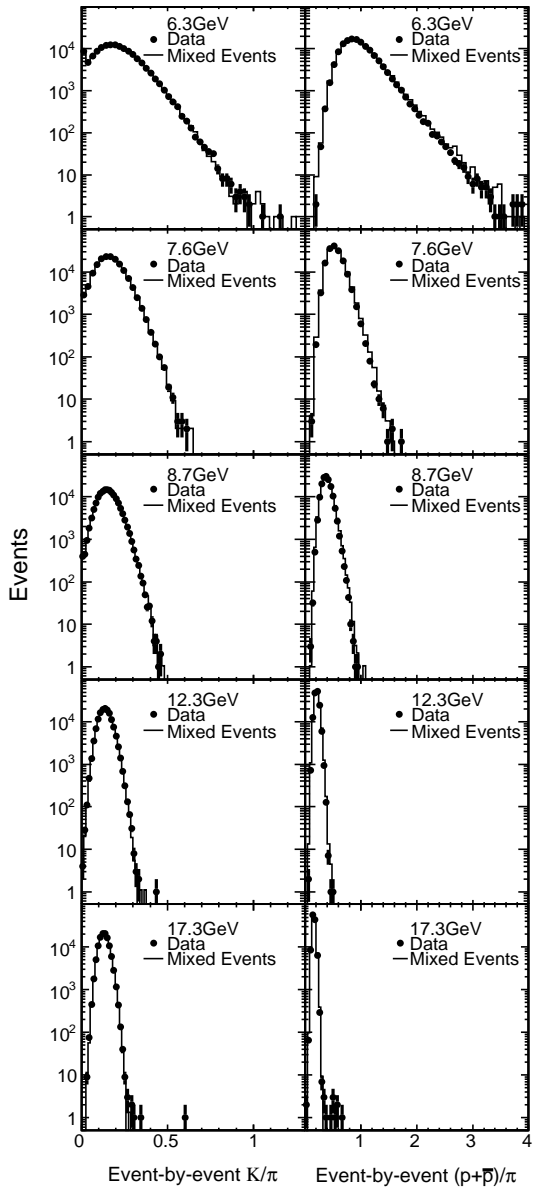}
\caption{\label{ratios} (Left) Distributions of the event-by-event \ktopi\ ratio 
for data (points) and mixed events (histogram), for all five collision energies. 
(Right) Distributions of the event-by-event \ptopi\ ratio 
for data (points) and mixed events (histogram), for all five collision energies.
No acceptance corrections were applied to the particle ratios shown.}
\vspace{-0.3cm}
\end{figure}

The results of the event-by-event fits for the five data sets are presented in Figure \ref{ratios}, showing the 
event-by-event ratio distributions for \ktopi\ (left) and \ptopi\ (right).
Points show the \ktopi\ and \ptopi\ distributions for data, while the histogram shows the corresponding reference distributions obtained for mixed events (see below).

\subsection{Measurement of non-statistical fluctuations}

The next step in the analysis is to extract the strength of non-statistical event-by-event fluctuations
from the observed distributions of \ktopi\ and \ptopi.  The relative width $\sigma$, defined as $\sigma = \mbox{RMS/Mean} \times 100~[\%]$, of the 
measured event-by-event particle ratio distributions can be decomposed into 
three contributions:
\begin{enumerate}
\item due to the finite number of particles produced and observed per event,
the ratio of particle multiplicities measured event-by-event will exhibit
statistical fluctuations with a width dictated by the individual particle
multiplicities within the acceptance,
\item due to non-ideal particle identification, these statistical fluctuations
will be smeared by the experimental \dedx\ resolution and the event-by-event
fitting procedure,
\item due to genuine non-statistical fluctuations, which if they exist are superimposed on the background of 
statistical and experimental fluctuations.
\end{enumerate}
The combined contributions of finite number fluctuations in the particle multiplicities and effects of detector resolution 
are estimated using a mixed event technique. Mixed events are constructed by 
randomly selecting measured particles from different events and combining them into artificial
events, while reproducing the multiplicity distribution of the real events. By construction, 
mixed events have on average the same particle ratios as the real events, but no internal 
correlations. In creating the mixed events, the \dedx\ measurement for each particle is carried
over to the mixed events. This allows us to estimate not only the finite-number fluctuations, but 
also the effect of the finite \dedx\ resolution using mixed events. The mixed events are 
subjected to the same fit procedure as the real events and the RMS-width $\sigma_{mix}$ 
of the mixed event particle ratio distribution is obtained. The results of the event-by-event 
fit on mixed events are compared in Figure \ref{ratios} to those from real
events. As the figure illustrates, the contributions from the statistical and experimental fluctuations dominate
the event-by-event particle ratio distributions, and have to be subtracted carefully.

To extract the physically relevant non-statistical fluctuations from the data distributions as shown in 
Figure \ref{ratios}, we calculate for each energy the RMS widths of the event-by-event ratio
distributions for data and mixed events. The non-statistical fluctuations $\sigma_{dyn}$ are then estimated 
by subtracting the relative RMS width of data and mixed event distributions in quadrature:
\begin{equation}
\sigma_{dyn} = sign(\sigma_{data}^2 - \sigma_{mixed}^2) \sqrt{|\sigma_{data}^2 - \sigma_{mixed}^2|}
\label{sdyn}
\end{equation}
It is important to note that non-statistical fluctuations $\sigma_{dyn} \neq 0$ can arise from correlated particle
production. Possible sources of particle correlations include energy-momentum and quantum-number conservation, 
quantum correlations and the decay of resonances (which includes strong and electromagnetic decays and a possible
contamination due to weak decays).  In particular, the definition of $\sigma_{dyn}$ allows for 
$\sigma_{data}^2 < \sigma_{mixed}^2$, which could result from correlated production of the 
two particle species forming the ratio, as expected for pion and proton production from hyperon decays.

\subsection{Systematic uncertainty estimates}
\label{systematics}

As the comparison of the event-by-event particle ratio distributions for data and mixed events shows, the overall width of the 
distributions is dominated by the fluctuations due to finite particle statistics and resolution.
Clearly, the measurement relies on the absence of additional detector-related fluctuations not accounted for by the
mixed-event technique and on the reliability of the event-by-event fitting procedure. Several tests on modified mixed events and
on input from transport model calculations were performed to verify these requirements. In addition, we varied the event 
and track selection criteria, as well as the parameters of the \dedx\ fits 
to estimate the associated systematic uncertainties. The results of these tests are described in the 
following section. Figures~\ref{linearity}--\ref{lowk2piratio} show the results for \ktopi\ ratios. 
For all tests, the uncertainties on \ptopi\ fluctuations were found to be smaller or equal to
those on \ktopi\ fluctuations. This is expected, as protons are farther separated in
\dedx\ from the dominant pion contribution. 

Figure~\ref{sys_error} summarizes the results of all of the systematic tests
described in this section.
The error bars in that figure show the estimated statistical 
uncertainty on the change of the \ktopi\ dynamical fluctuations for each variation of 
fitting procedure or track and event selection. For tests performed on data, as discussed in Sections \ref{systematics}~3 and~6, 
MC estimates of the correlation coefficient for the fit results on different samples were used to take 
into account the large overlap between the samples when e.g.\ removing the top or bottom 1\% of events or 
changing track quality cuts.

\subsubsection{Linearity of event-by-event fit}

Using the PDFs as a function of ($\vec{p},dE/dx$) extracted from the event-averaged data, the particles in 
the mixed event pool can be assigned a particle species, such that on average they conform to the known PDFs. This assignment 
then allows us to create modified mixed events with the desired multiplicity distribution and 
adjustable ratios of kaons, pions and protons.  Thus we are able to vary both the 
mean and fluctuations in e.g.\ the \ktopi\ ratio in the modified mixed events. Using this technique, we created modified mixed events with
\ktopi\ ratios varying from 0.08 to 0.26 using the multiplicity distribution for $\roots = 17.3$~GeV and subjected these 
mixed events to the event-by-event fit. As shown in Fig.\ref{linearity}, this test demonstrated that over a wide range 
in \ktopi\, the event-by-event ratio estimator is linear with respect to the true \ktopi\ ratio in the event. A 
small bias of 0.002 in the extracted particle ratio is observed. As the same fitting procedure is used on data and mixed
events, this bias will not affect the estimate of $\sigma_{dyn}$.

\subsubsection{Sensitivity of fluctuation measurement}

\begin{figure}[htp]
\begin{minipage}[t]{0.45\linewidth}
\centering
\includegraphics[width=4.25cm,clip]{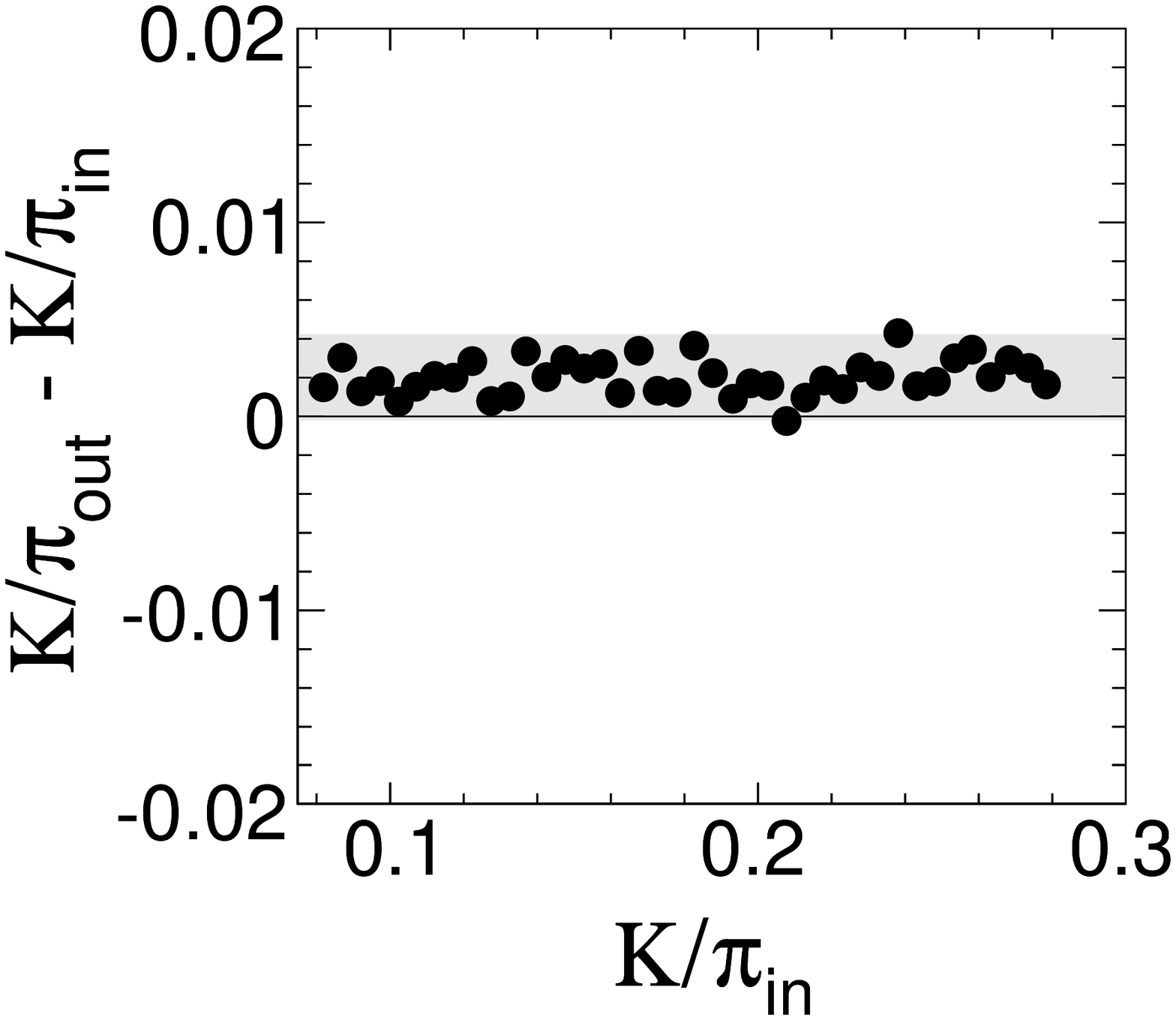}
\caption{\label{linearity} Difference between measured and input \ktopi\ ratio versus average 
input \ktopi\ ratio for mixed events, showing that 
the event-by-event fit is linear with respect to the input ratio.}
\end{minipage}
\hspace{0.2cm}
\begin{minipage}[t]{0.45\linewidth}
\centering
\includegraphics[width=4.25cm,clip]{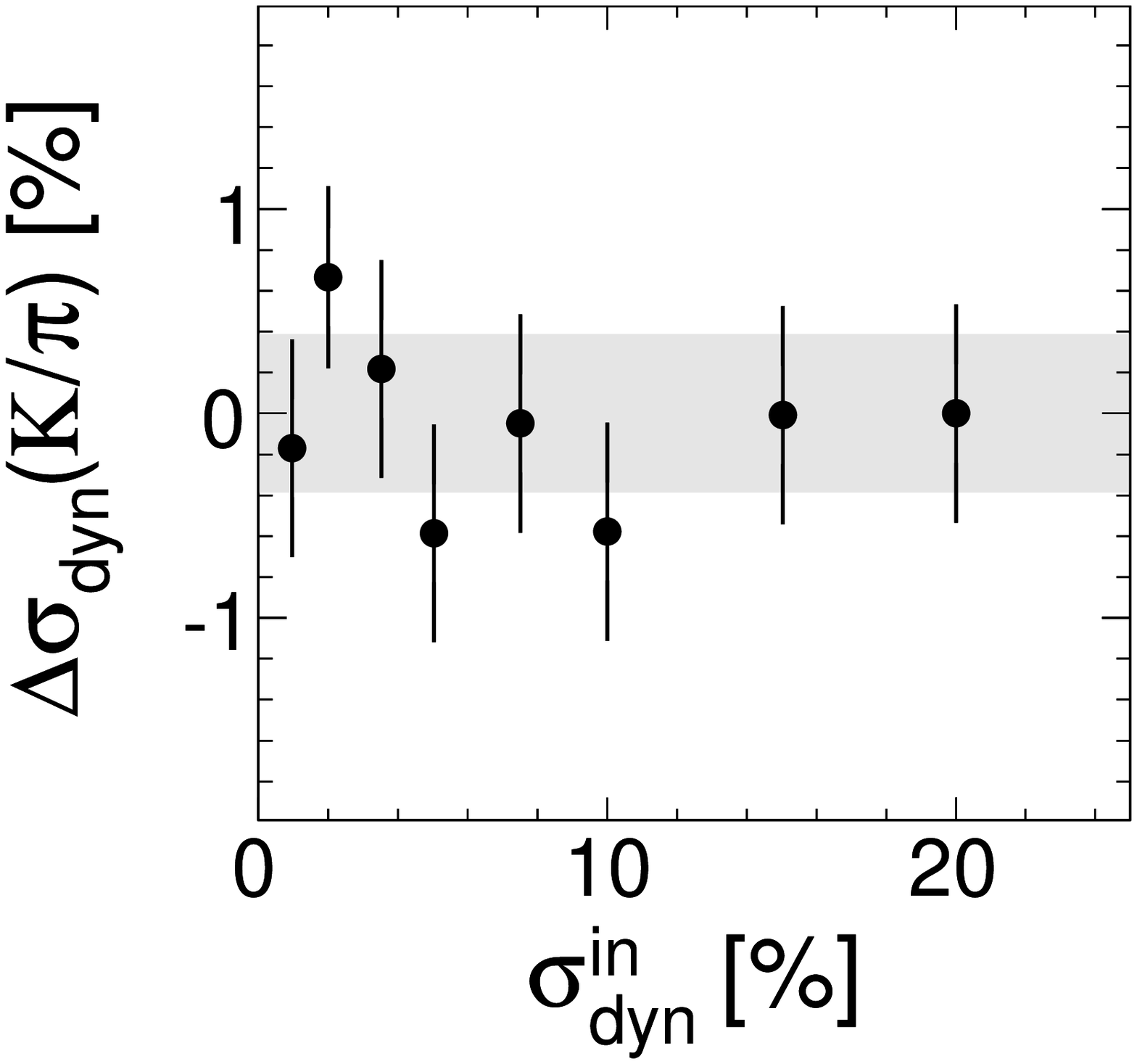}
\caption{\label{sensitivity} Relative deviation of extracted \ktopi\ non-statistical fluctuations
from the input value, $\sigma_{dyn}^{in}$, as a function of $\sigma_{dyn}^{in}$
for modified mixed events (see text). The shaded band indicates the 
assigned estimate of systematic uncertainty from the fitting procedure as a function of amplitude.}
\end{minipage}
\end{figure}

The modified mixed events discussed in the previous section also allow us to test the sensitivity of the extraction
of non-statistical ratio fluctuations. Using the multiplicity distribution of the 17.3 GeV dataset, non-statistical Gaussian variations
of magnitude $\sigma_{dyn}^{in}$  in the \ktopi\ ratio were introduced as a ``signal'' in the creation of the modified mixed events, 
in addition to the statistical event-by-event variation. For values 
of $\sigma_{dyn}^{in}$ from 0 to 20\%, these modified mixed events were then run through the full analysis procedure, using unmodified 
mixed events as the statistical reference. The result of this study is shown in Figure \ref{sensitivity}, where the extracted 
fluctuations are compared to the known input amplitude. The extracted magnitude of fluctuations is found to be equal to the 
input amplitude within statistical errors, even for small input fluctuations of $\approx 1$\%. A small systematic uncertainty
of 0.4\% is assigned to the fitting procedure for both \ktopi\ and \ptopi\ non-statistical fluctuations.

Another concern is the highly non-Gaussian nature of the event-by-event \ktopi\ distributions for low particle multiplicities. 
To check the validity of equation~2 in this case, the observed distributions for mixed events were fitted with Gamma-distributions
at each energy. The Gamma-distributions were folded with Gaussians corresponding to the magnitude of non-statistical fluctuations 
at each energy and sampled to reproduce the input histograms for $\sigma^2_{data}$. The unfolded Gamma-distributions were sampled
to obtain $\sigma^2_{mixed}$. The resulting $\sigma_{dyn}$ was plotted for each energy, in comparison to the corresponding 
input value, in Figure \ref{sys_error}(a). This study demonstrates that the magnitude of the Gaussian input fluctuations
is recovered from applying equation~2 to the RMS width of the non-Gaussian signal and reference ratio distributions.
The effect of low particle multiplicities was also tested by repeating the analysis for the highest energy  dataset
while randomly removing particles from each event, to resemble the multiplicity distributions in the lower energy
datasets. The extracted particle ratios were found to be independent of the fraction of removed particles within statistical 
errors.

Finally, the analysis method was tested on events simulated with the UrQMD transport code \cite{urqmd}. For these events,
non-statistical fluctuations as a function of beam-energy were determined using two methods. The first method determined the 
event-by-event particle ratios by counting the different particle multiplicities in the NA49 acceptance. The second method
assigned a \dedx\ value to each particle according to the ($\vec{p}, dE/dx$) PDFs and then performed an event-by-event fit to extract
the particle ratios. For each method, the non-statistical fluctuations were extracted by comparison to the respective mixed event 
reference. As Figure \ref{sys_error}(b) demonstrates, the two methods yield the same magnitude of non-statistical fluctuations at 
each collision energy, within statistical errors. To account for a possible small energy dependence, the systematic uncertainty shown by
the shaded area in Figure \ref{sys_error}(b) was assigned.

\subsubsection{Fluctuations in the tails of ratio distributions}
\label{tails}

In Figure~\ref{ratios}, several outlier events in the ratio distributions can be seen, in particular for the higher
collision energies. To test whether the observed non-statistical fluctuations are due to these outliers or
fluctuations in the tails of the ratio distributions, we removed 1\% of events with the highest
\ktopi\ ratio in data and mixed events at each energy. The resulting change of the extracted fluctuation
signal is shown by the full symbols in Figure~\ref{sys_error}(c). Also plotted in this figure is the result
of removing the 1\% of events with the lowest \ktopi\ ratio.
As expected, removing either of the tails of the distribution leads to
a slight reduction in the extracted dynamical fluctuations $\sigma_{dyn}$. In particular for removing the high tail
of the distribution, the change is small for most energies. Removing the low end of the distribution
has a 10\% effect at the low energies. Smaller changes were seen for a corresponding
study of \ptopi\ fluctuations.

\begin{figure}[htp]
\begin{minipage}[t]{0.45\linewidth}
\centering
\includegraphics[width=4.25cm,clip]{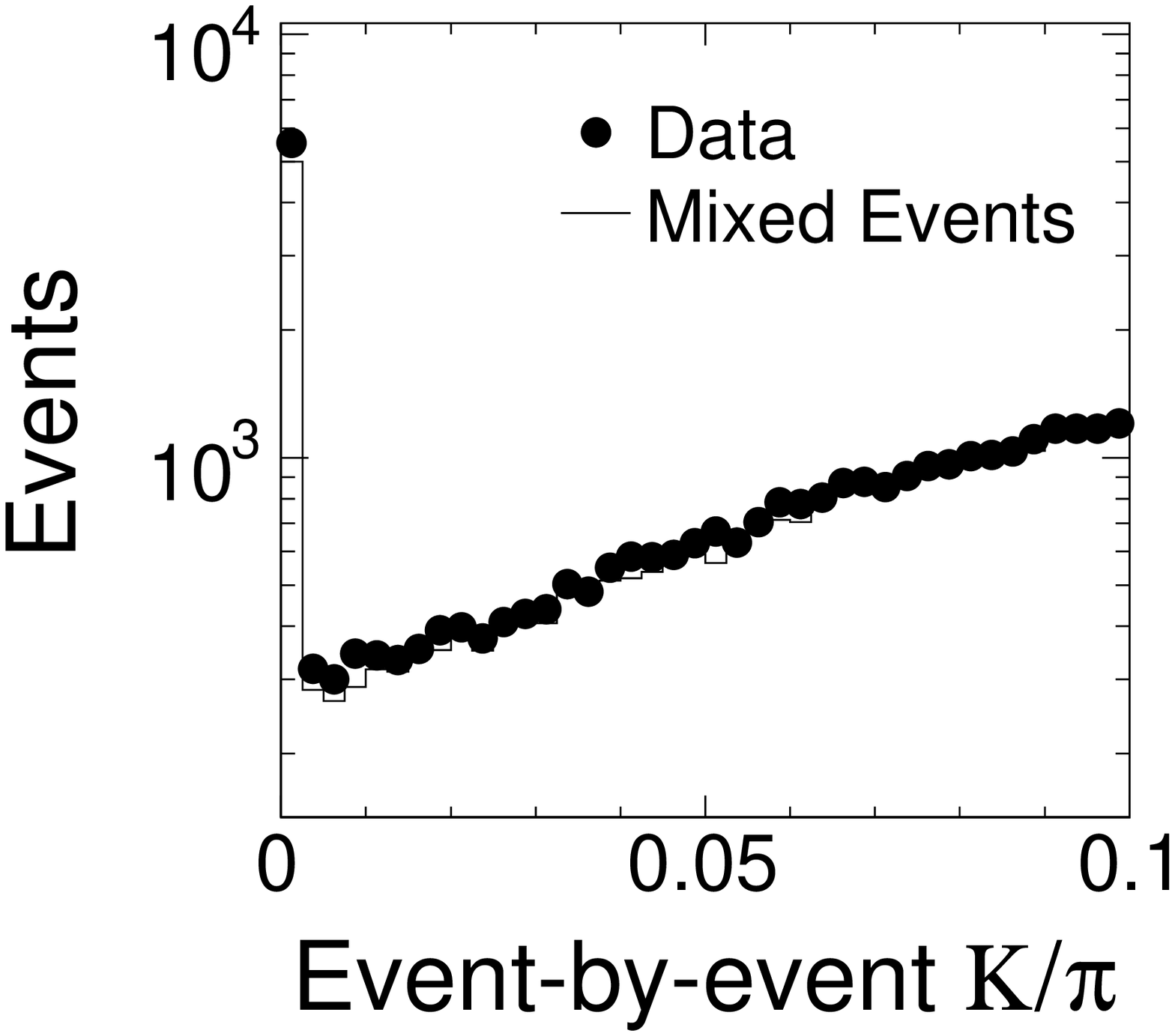}
\caption{\label{lowk2pizoom} Expanded view of the low \ktopi\ ratio region of the event-by-event \ktopi\ ratio distribution of the $\roots = 6.3$~GeV data set for data (points) and mixed events (histogram).}
\end{minipage}
\hspace{0.2cm}
\begin{minipage}[t]{0.45\linewidth}
\centering
\includegraphics[width=4.25cm,clip]{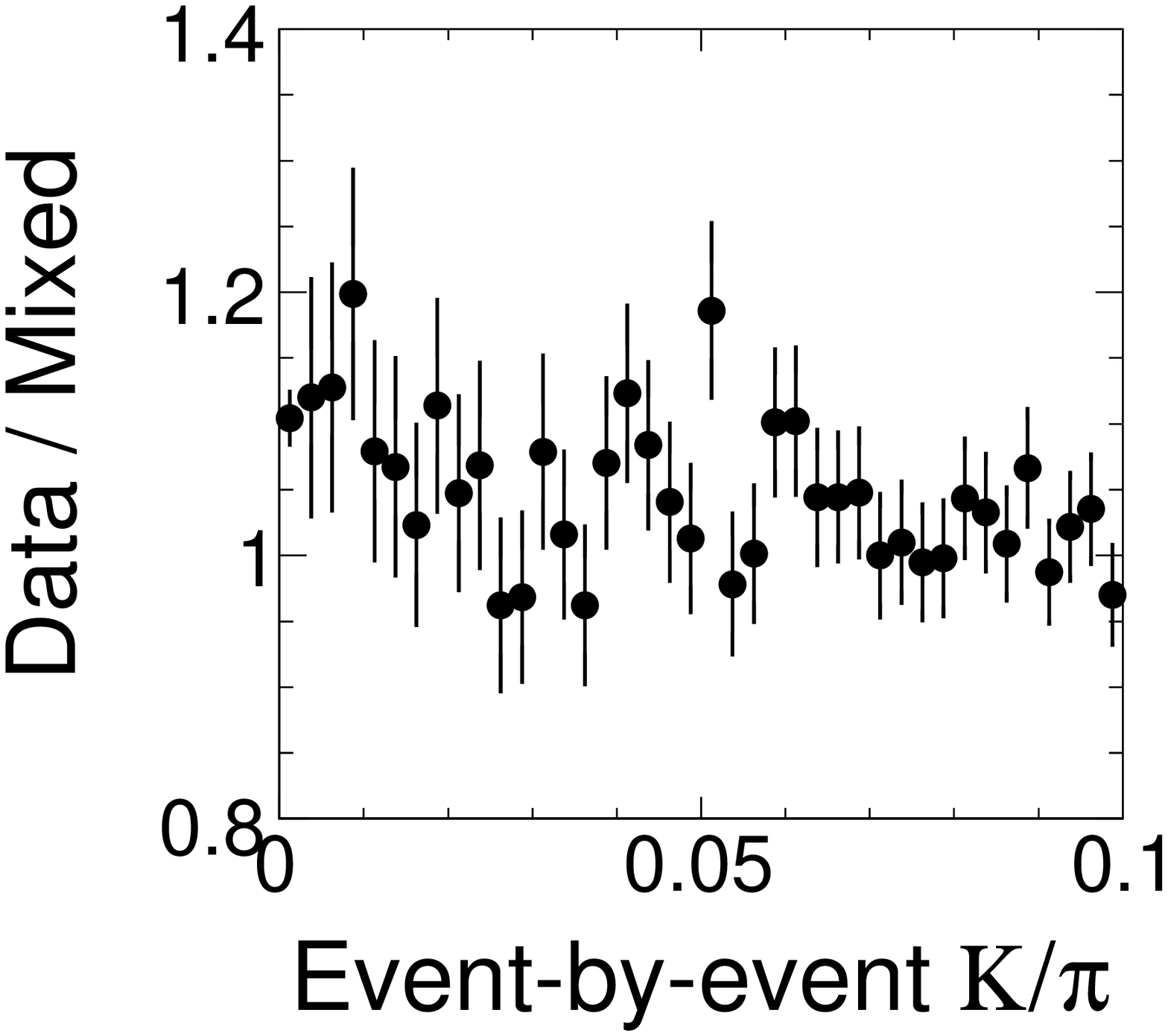}
\caption{\label{lowk2piratio} Ratio of the event-by-event \ktopi\ ratio distribution for data and mixed events of the $\roots = 6.3$~GeV data set.}
\end{minipage}
\end{figure}

The \ktopi\ ratio distributions for the lowest two energies develop a spike at very small \ktopi\ ratios. 
Figure~\ref{lowk2pizoom} shows an expanded view of the small \ktopi\ ratio region for the $\roots = 6.3$~GeV 
data set. We examined the events in this region for signs of detector or
reconstruction failures, but found no anomalies compared to the bulk of events. The spike is also present
in the mixed event sample, suggesting that it arises from a combination of low kaon multiplicities 
and the finite resolution of the event-by-event maximum likelihood fit. It is important to note that the fit
does not allow negative \ktopi\ values. To confirm this hypothesis, we studied the ratio
of the event-by-event \ktopi\ distributions in data and mixed events, shown in Figure~\ref{lowk2piratio}. 
This ratio does not exhibit a spike at small \ktopi\ ratios. 
The shape of this ratio is consistent with the presence of dynamical fluctuations in the data events and
with the observed small decrease in $\sigma_{dyn}$ when removing events in the low tails of the
distributions from the analysis.

Finally, we performed further simulations using the UrQMD generator. For
generated UrQMD events, event-by-event \ktopi\ distributions were obtained by counting particles in the NA49 acceptance
and alternatively by assigning energy loss information to each particle and performing the same event-by-event likelihood
fit as described before. As expected, a broadening of the event-by-event \ktopi\ distribution is seen when using the likelihood
fit as compared to particle counting. This leads to an enhanced number of events in the $\ktopi\ \approx 0$ bin, reproducing
what is seen in the data.
However, this enhancement is again well reproduced in the corresponding mixed events and no bias is found
compared to an analysis of dynamical fluctuations based on direct counting of particles.

In summary, our studies show that the extreme tails of the distribution do not 
dominate the overall fluctuation strength. In particular the shape of the \ktopi\ distributions 
for small \ktopi\ is well reproduced by the mixed event procedure and in UrQMD studies using simulated 
energy loss information. 

\subsubsection{Uncertainty in \dedx\ scaling}

In determining the \dedx\ PDFs, we assume that the width of the \dedx\ distributions for the different
particle species scales with the average \dedx\ as $\dedx^\alpha$, with $\alpha = 0.65$ obtained in
previous studies of \dedx\ particle identification \cite{MarcoThesis}. As a function of momentum $\vec{p}$, 
the width for a given particle species also changes due to variations in the observed TPC track length
and due to variations in the local track density. In combination with the small separation between 
different particle species, this makes a precise determination of $\alpha$ difficult. To test the 
effect of the uncertainty in $\alpha$ on the observed fluctuations, we repeated the analysis, starting from
the determination of PDFs, for a much larger value of $\alpha = 1.15$. The difference between the fluctuation
results obtained for $\alpha = 1.15$ and $\alpha = 0.65$ is shown in Figure~\ref{sys_error}(d). The fluctuations
are seen to be rather robust against changes in $\alpha$, with the difference only reaching up
to 1\% for the lower collision energies. Even smaller variations are seen in the \ptopi\ 
fluctuations. The actual uncertainty in $\alpha$ is estimated to be only $\pm 0.15$, 
leading to the systematic uncertainty estimate shown as the shaded band in Figure~\ref{sys_error}(d).

\subsubsection{Track quality cuts}
\label{track_quali}

In order to test for possible systematic distortions of the fluctuation measurement not included in the MC simulations,
we used two sets of quality cuts in the event and track selection for the analysis. 
The corresponding datasets should have different sensitivity to detector and tracking effects,
such as contributions from weak decays and secondary interactions, or effects from varying 
track densities on reconstruction and particle identification.
The quality cuts are based on the distance of closest approach of the extrapolated particle 
trajectory to the main vertex, the length over which the track was measured in the TPC and the number of measured
points on the track. For the set of loose quality cuts, the cuts 
on the extrapolation of tracks to the primary vertex and a cut on the track fit quality were removed.
As shown in Table~1, the difference in multiplicity for the two samples varies with increasing track 
density and changing acceptance, as the collision energy increases, and ranges from 10\% at the lowest energy to about 20\% at the highest energy.
The track selection using loose track quality cuts yields more tracks per event to include in the event-by-event 
particle ratio estimation,  but possibly also increases the contamination of the track sample with non-primary and background 
particles. 

\begin{figure}[t]
\begin{center}
\includegraphics[width=8.5cm]{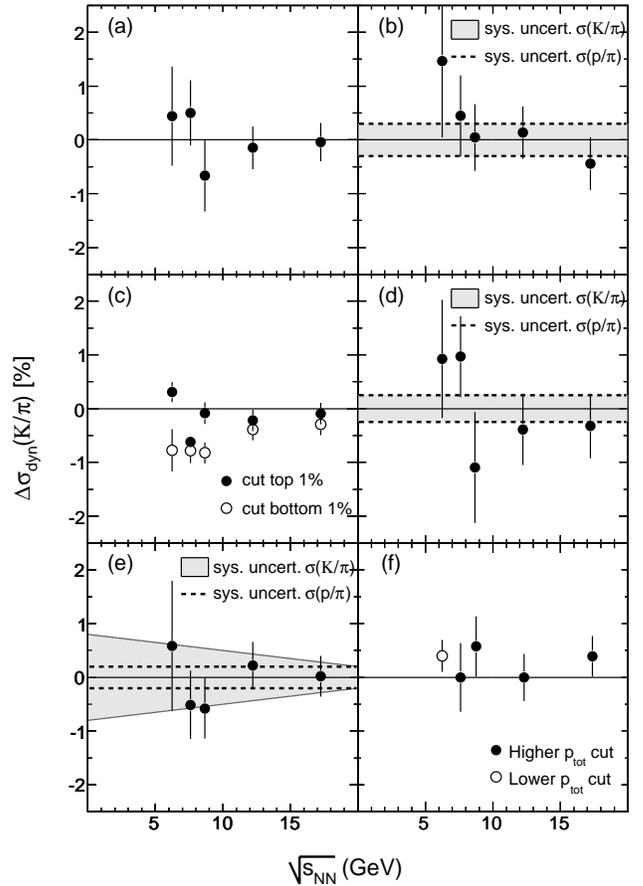}
\end{center}
\caption{\label{sys_error} Results of various studies on systematic uncertainties in the determination of $\sigma_{dyn}(K/\pi)$:
(a) extraction of $\sigma_{dyn}$ using equation~2. (b) \dedx\ fit versus particle counting in UrQMD. (c) removal 
of tails of event-by-event distributions. (d) variation of \dedx\ width scaling. (e) variation of track
quality cuts. (f) variation of $p_{tot}$ cutoff. Shaded bands indicate the contribution to the systematic error
for \ktopi\ fluctuations. Where applicable, dashed lines indicate corresponding contribution to systematic error for 
\ptopi\ fluctuations. See text for details.
}
\end{figure}

The final value of the fluctuation signal presented below is calculated as the arithmetic mean of the results from both
samples. Figure~\ref{sys_error}(e) shows the deviation from the average result for the tight track-cut sample. 
The shaded area indicates
the estimated systematic uncertainty in the final result for \ktopi\ fluctuations related to the track selection. The same study
was performed for \ptopi\ fluctuations, resulting in a smaller variation than seen for \ktopi. The corresponding contribution to 
the systematic uncertainty on the \ptopi\ non-statistical fluctuations is shown by the dashed lines in Figure ~\ref{sys_error}(e).

\subsubsection{Effects of acceptance variation}
\label{accvar}

As discussed above, the constant PID cutoff in lab total momentum $p_{tot} > 3$~GeV/c, below which particles species
cannot reliably be separated, leads to a significantly changing acceptance at different
beam energies (see Fig.~2). To study the influence of the changing acceptance relative to the lowest beam
energy, we modified the cutoff for the four higher beam energies to 3.6, 4.4, 6.3 and  9.1~GeV/c, respectively.
These higher cutoffs remove low $p_T$ particles near midrapidity approximating
the default cutoff for the $\roots = 6.3$~GeV data sample. The resulting change in the \ktopi\ fluctuation signal is 
plotted as the filled symbols in Figure~\ref{sys_error}(f), showing a rather small difference compared to the standard acceptance cut. 
We also studied a reduced $p_{tot}$ cut of 1.45~GeV/c for the lowest energy data set. 
The region $1.45~\mbox{GeV/c} < p_{tot} < 3~\mbox{GeV/c}$ yields little 
separation between kaons and protons, but extends the pion acceptance closer to mid-rapidity. 
The corresponding fluctuation signal is shown as the open symbol in Figure~\ref{sys_error}(f). 
Again, no strong change in the \ktopi\ non-statistical fluctuations is observed.  
Note that the event multiplicity is changed by up to 30\% by the modified cuts, such that part of the 
change observed is likely due to statistical fluctuations. As any remaining variation could be due to the
physical nature of the non-statistical fluctuations, no systematic uncertainty is assigned for the acceptance
variations.

While this study indicated that the results for \ktopi\ fluctuations only have a moderate dependence on the details 
of the acceptance, we need to stress again that a detailed comparison to theoretical models needs to take the 
experimental acceptance into account. As the measurement is performed within a limited acceptance,
particle ratio fluctuations within the acceptance can in principle be caused by non-statistical 
variations of e.g.\ the kaon phase space distribution, rather than a non-statistical variation of the \ktopi\ 
ratio in full phase space. Note however that purely statistical variations of the phase-space distributions
are removed by comparison to the mixed event reference.

\section{Results and discussion}

The event-by-event \ktopi\ and \ptopi\ ratios shown in Figure \ref{ratios}
exhibit smooth distributions at all five collision energies and show no evidence 
for the existence of distinct event classes. Qualitatively the \ktopi\ ratio
distributions are found to be wider than the statistical reference for 
all energies, while the 
\ptopi\ ratio distributions are narrower than the respective reference.

\begin{figure}[t]
\includegraphics[width=9cm]{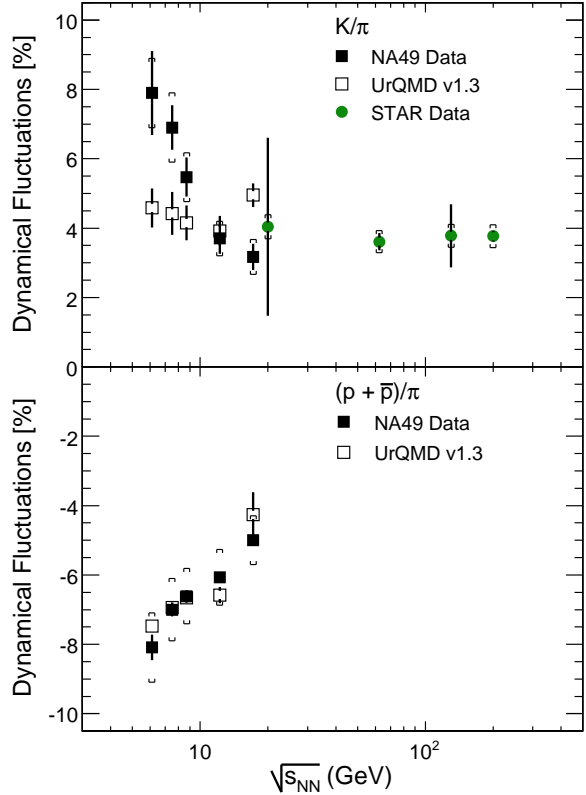}
\caption{\label{edep} (Color online) Energy dependence of the event-by-event non-statistical fluctuations of the \ktopi\ ratio
(top panel) and the \ptopi\ ratio (bottom panel). Filled symbols show data, open symbols show calculations
with the UrQMD transport code, using NA49 acceptance tables.  Systematic uncertainties are shown as brackets. 
Data of the STAR collaboration are taken from \cite{star_k2pi}.}
\end{figure}

The deviation from the statistical reference $\sigma_{dyn}$, as defined in equation \ref{sdyn}, 
is plotted in Figure \ref{edep} for 
\ktopi\ fluctuations (top) and \ptopi\ fluctuations (bottom). As stated in Section \ref{track_quali}, the final
data represent the average of the loose and tight track quality cut results.
The observed non-statistical fluctuations of the \ktopi\ ratio are positive and decrease with collision energy. For the 
highest collision energy, a value of $\sigma_{dyn} = 3.2 \pm 0.4 \mbox{(stat.)} \pm 0.5 \mbox{(syst.)} \%$ 
is observed, in agreement with the results
reported in \cite{oldk2pi} for an independent dataset. 
Towards lower collision energies, a steep increase of the fluctuation signal is observed,
the \ktopi\ fluctuations reaching a value of 
$\sigma_{dyn} = 7.9 \pm 1.2 \mbox{(stat.)} \pm 1.0 \mbox{(syst.)}\% $ for the lowest energy.
The increase is qualitatively consistent with calculations assuming the presence of
a deconfinement phase transition at the lowest collision energies of the CERN SPS~\cite{Gorenstein:2003hk}. 
This scenario predicts small and energy-independent fluctuations in the \ktopi\ ratio in the QGP phase, while
for the confined phase large fluctuations are expected that increase towards lower collision energies.

For the \ptopi\ ratio, which has not been previously reported,
the non-statistical fluctuation measure $\sigma_{dyn}$ is negative for all collision energies. For the lowest
collision energy a value of $\sigma_{dyn} = -8.1 \pm 0.4 \mbox{(stat.)} \pm 1.0 \mbox{(syst.)}\% $
is observed. The negative value of $\sigma_{dyn}$ for \ptopi\ fluctuations  reflects the fact that the 
width of the data distribution is smaller than the width of the distribution of mixed events for 
this ratio.
A negative fluctuation signal of the event-by-event \ptopi\ ratio can result
from resonance decays into pions and protons. Such contributions correlate the 
pion and proton multiplicities event-by-event, and thus lead to smaller fluctuations in the ratio
than expected from independent particle production.
The magnitude of the negative fluctuation signal in the \ptopi\ channel is then
related to the relative contribution of resonance decay products in the final state of the collision.

In order to further estimate the influence of known correlations in the events, like those induced by 
conservation laws and particle production from resonance decays, we 
analyzed fluctuations of the \ktopi\ and \ptopi\ ratios in a string-hadronic cascade model, UrQMD \cite{urqmd}. 
In this model, by construction, no fluctuations due to a potential phase transition are present. 
For this study, large samples of UrQMD events were generated at all five collision energies and then subjected 
to an acceptance filter modeling the NA49 detector system. The accepted final state particles 
were counted and the corresponding ratios were formed for each event.
The energy dependence of the event-by-event \ptopi\ ratio fluctuations in UrQMD closely matches the 
energy dependence observed in the data, as shown in the bottom plot in Figure \ref{edep}. This lends further support to 
interpreting the negative fluctuation signal as resulting from resonance decays, which are the dominant
source of \ptopi\ fluctuations in UrQMD.

The energy dependence of the \ktopi\ ratio fluctuation signal however is not reproduced in the cascade model,
which gives an energy independent fluctuation signal. Further studies are needed to determine 
whether  the finite non-statistical \ktopi\ fluctuations in UrQMD originate from resonance decays or 
correlated particle production due to conservation laws.

The \ktopi\ fluctuation signal in the data taken at $\roots = 17.3$~GeV collision 
energy was found to be consistent with calculations performed assuming a grand canonical ensemble 
without enforcing local conservation laws \cite{koch99}. The small fluctuations at this energy are 
thus consistent with every event being a random sample from the identical thermal ensemble. 
The interpretation of the observed increase of \ktopi\ fluctuations towards lower energies is complicated by the 
simultaneous decrease in true particle multiplicities towards lower energies. Stephanov has 
pointed out that the observed energy dependence could approximately be understood as the combination of a 
fixed pair-wise correlation strength, in combination with the known dependence of the average event multiplicity
on collision energy \cite{StephThis}. Within the experimental resolution of the present measurement alone, the 
multiplicity scaling expected from this argument can not be ruled out.

Recently, the STAR collaboration has reported preliminary results for non-statistical \ktopi\ fluctuations in central 
Au+Au collisions in the RHIC energy range from $\roots = 19.6$ to 200~GeV \cite{star_k2pi}.
The reported results for 19.6~GeV are in agreement with the values quoted here for $\roots = 17.3$~GeV, within statistical
errors. Although a detailed quantitative comparison of the results suffers from the differences in the NA49 and STAR acceptances, it is
important to note that the STAR data show no significant energy dependence, in contrast to the results reported here
for the SPS energy range. If confirmed, this would rule out, or at least limit the range of validity, for the 
proposed multiplicity scaling discussed above.

\section{Summary}
We have presented a study of the energy dependence of non-statistical event-by-event fluctuations of the 
\ktopi\ and \ptopi\ ratios in the energy range $6.3 < \roots < 17.3$~GeV. 
A strong increase of the \ktopi\ fluctuation signal is observed towards the low end 
of this energy range. 
The increase of the signal is not reproduced by 
the UrQMD hadronic cascade model, suggesting the onset of a new source of fluctuations. 
The same model is however able to describe the strong fluctuations seen in the \ptopi\ ratio.
The domain of increased \ktopi\ fluctuations coincides with the energy range where a change in the 
behavior of event-averaged hadronic observables in heavy-ion collisions occurs, relative to elementary collisions.  
Further theoretical calculations will be needed to evaluate the relevance of this observation for a possible 
interpretation in the context of a deconfinement phase transition.

This work was supported by the US Department of Energy Grant DE-FG03-97ER41020/A000,
the Bundesministerium fur Bildung und Forschung, Germany, 
the Polish State Committee for Scientific Research (1 P03B 006 30, 1 P03B 097 29,
1 P03B 121 29, SPB/CERN/P-03/Dz 446/2002-2004, 
2 P03B 04123), 
the Hungarian Scientific Research Foundation (T032648, T032293, T043514),
the Hungarian National Science Foundation, OTKA, (F034707),
the Polish-German Foundation, and the Korea Science \& Engineering Foundation Grant (R01-2005-000-10334-0)

\end{document}